\documentclass[12pt]{article}
\usepackage{epsfig}
\def\la{\langle}
\def\ra{\rangle}
\def\pa{\partial}

\def\tninth{{\textstyle\frac{1}{9}}}
\def\tthird{{\textstyle\frac{1}{3}}}

\def\half{\frac{1}{2}}
\def\sixth{\frac{1}{6}} 

\def\third{\frac{1}{3}}

\def\tansq{\la{{\bf X'}^2\ra}}
\def\velsq{\la{\dot {\bf X}^2\ra}}
\def\vsq{\bar{v}^2}
\def\tsq{\bar{t}^2}

\def\psdel{\la {|\Delta_a(k,\eta)|^2\ra}}
\def\pseng{\la {|\Theta_{00}(k)|^2\ra}}
\def\psmom{\la {| U(k)|^2\ra}}
\def\pscross{\la {U(k)\Theta^*_{00}(k)\ra}}

\def\psengt{\la {|\Theta_{00}(k,\eta)|^2\ra}}
\def\psmomt{\la {| U(k,\eta)|^2\ra}}

\def\TTeng{\la {\Theta_{00}(k,\eta)\Theta_{00}^*(k,\eta')\ra}}
\def\TTmom{\la {U(k,\eta)U^*(k,\eta')\ra}}
\def\TTcross{\la {U(k,\eta)\Theta_{00}^*(k,\eta')\ra}}

\newcommand\ben{\begin{equation}}
\newcommand\een{\end{equation}}
\newcommand\bea{\begin{eqnarray}}
\newcommand\eea{\end{eqnarray}}

\title{
\vspace{-36pt}
{\normalsize \begin{flushright}SUSX-TH-96-008\\
{\tt astro-ph/9606137}\\
\end{flushright}}
\vspace{1 cm}
{\bf Correlations in Cosmic String Networks}}
\author{
Graham R. Vincent$^{(a)}$\thanks{E-mail: {\tt 
g.r.vincent@sussex.ac.uk}}\\
Mark Hindmarsh$^{(a)}$\thanks{E-mail: {\tt 
m.b.hindmarsh@sussex.ac.uk}}\\
Mairi Sakellariadou$^{(b)}$\thanks{E-mail: {\tt 
mairi@karystos.unige.ch}}}
\date{\today}

\begin{document}

\maketitle
\vspace{-30pt}
\begin{center}
{\normalsize {\it
$^{(a)}$School of Mathematical and Physical Sciences\\ 
University of Sussex\\
Brighton BN1 9QH\\
UK\\[6pt]
$^{(b)}$D\'epartement de Physique Th\'eorique\\
Universit\'e de Gen\`eve\\
Quai Ernest-Amsermet 24\\
CH-1211 Gen\`eve 4\\
Switzerland}
}
\end{center}

\vfill

\begin{abstract}
We investigate scaling and correlations of the energy and momentum in
an evolving network of cosmic strings in Minkowski space. These
quantities are of great interest, as they must be
understood before accurate predictions for the
power spectra of the perturbations in the matter and radiation
in the early Universe can be made. We argue that Minkowski space
provides a reasonable approximation to a Friedmann
background for string dynamics and we use
our results to construct a simple model of the network, in which it is
considered to consist of randomly placed
segments moving with random velocities. This model works well
in accounting for features of the two-time correlation functions,
and even better for the power spectra.
\end{abstract}
\newpage

\section{Introduction}

Phase-ordering dynamics after a cosmological phase transition
may explain the origin of structure in our universe \cite {ShelVil}.
Our interest is in the role of one class of theories - cosmic strings
\cite {HindKib}.

Recent attention has focused on cosmic strings formed during the 
breaking of
the grand unified symmetry, which naturally generates perturbations
in the matter and radiation fields of the right order for structure
formation. However, to 
make diagnostic predictions between cosmic strings, other 
defects and inflation-based theories we need to use a detailed theory
of the cosmological evolution of perturbations.
There exists a well-founded theory
for the standard inflationary scenario of gaussian fluctuations
in the gravitational potential. In extending the theory to cover 
defect-based
theories there are three principal issues to be faced.
Firstly, although the initial conditions are random, they are not
gaussian which makes averaging over the ensembles of defects a 
numerical problem. Secondly, the creation
of the defects must conserve energy-momentum: hence there are 
compensating fluctuations
in the fluid components on super-horizon scales.
Thirdly, in the subsequent evolution, the defect stress-energy 
$\Theta_{\mu\nu}$ must
be included in the Einstein equations. Defects form an unusual 
component of the
total stress-energy, in that the metric perturbations
only affect the evolution of the defect network to second order, so in 
the linear
approximation we can evolve the network in the unperturbed background 
metric. Its stress-energy then acts as a source term for the familiar
set of differential equations governing the evolution of the 
perturbations.
As demonstrated by Veeraraghavan and Stebbins \cite {VeeSte90}, the 
solutions to the Einstein equations 
describing the {\em subsequent} evolution of perturbation variables 
with
a source term can be expressed as a convolution of a suitable Green's 
function
(dependent on the usual cosmological parameters) with the source term 
integrated
over time. For a set of perturbation variables $\Delta_a$ and a set of 
source
terms $S_b$ related to $\Theta_{\mu\nu}$
\ben
\Delta_a(k,\eta)=\int_{\eta_i}^\eta d{\eta}'\,G_{ab}(\eta,\eta') 
S_b(k,\eta')
\label{pert1}
\een
where $\eta$ and $\eta'$ are conformal times and {\em k} is the 
wavenumber of the mode.
Calculating the power in these variables will then involve integrating
over the two-time correlator
\ben
\psdel=\int_{\eta_i}^{\eta} d{\eta}'\,d{\eta}''\,G_{ab}(\eta,\eta') 
G_{ac}(\eta,\eta'') \la{S_b(k,\eta') S_c(k,\eta'')\ra}
\label{pert2}
\een
When using the Veeraraghavan-Stebbins formalism, the Green's functions 
can be calculated
from standard cosmological perturbation theory:
analytically in the simplest cases, or numerically for a more realistic 
universe.
However, in the absence of a workable analytic treatment of string 
evolution
(although see \cite {AusCopKib} and \cite{Hind1})
we must rely for the moment on a numerical study of the two-time 
correlators.

In a simplified model the Universe consists of two fluids: a 
pressureless CDM component
and a tightly-coupled baryon-photon fluid. In the convenient 
synchronous gauge, the sources are $\Theta_{00}$ and $i\hat 
k_j\Theta_{0j}$ (which we 
shall hereafter call $U$), when the equations are rewritten
using the energy-momentum pseudotensor introduced by Veeraraghavan and
Stebbins \cite {VeeSte90}. In the Newtonian gauge they are also the 
sources
of curvature fluctuations \cite{HuSpeWhi96}.
It is therefore the two-time correlation
functions of these variables that we study.
Such a study is particularly timely in view of recent work
on the power spectrum of Cosmic Microwave Background around the degree 
scale from
strings \cite {MACF1,MACF2} and from global textures \cite 
{DuGaSa,CritTur}.
As emphasised by Albrecht {\em et\, al\,} \cite {ACFM}, the distinctive 
appearance
of ``Doppler'' peaks and troughs seen in inflationary calculations and 
texture
models depend sensitively on the temporal coherence of the sources. If 
one assumes little coherence the peaks are washed out; on the other 
hand
an assumption of total coherence preserves them. Magueijo {\em et\, 
al\,}
in \cite {MACF1} assumed that strings were 
effectively incoherent and obtained a rather featureless CMB power 
spectrum
at large multipole $l$. In \cite {MACF2} this assumption was justified
by a numerical study of the two-time energy-density correlator.

In contrast both Durrer {\em et\, al\,} \cite {DuGaSa} and  Crittenden
and Turok \cite {CritTur} assumed maximum coherence for their texture 
models and found
that the peaks were preserved, albeit in positions typical of 
isocurvature
models \cite{HuSpeWhi96}. It is therefore clear that understanding
the temporal coherence of string sources is of great importance when 
calculating their microwave background signals.

In this paper we present the 
results of some numerical ``experiments'' evolving strings in
Minkowski space. Using Minkowski space rather than a Friedmann model 
may seem
rather unrealistic. However, we expect a network of strings evolving in
Minkowski space to have all the essential features of one in a
Friedmann background. A tangled initial state consisting of a few
probably ``infinite'' strings plus a scale free distribution of loops
straightens out under tension and continually self intersects resulting
in the transfer of energy into very small fast moving loops. The 
infinite
string approaches ``scaling'' in which it is characterised as a set
of more or less Brownian random walks. The step length of the walks
and their average separation are both approximately equal to an overall
network scale $\xi$, which increases linearly with time (as naive
dimensional analysis would suggest). 
The length density of the infinite string
decreases as $\xi^{-2}$, again as dimensional analysis suggests.
In fact, $\xi$ is conventionally defined so that the length density
is precisely $1/\xi^2$ \cite {CopKibAus}.
The effect of an expanding background is to damp coherent motions on
super-horizon scales. However the network scale is much less that the 
horizon
size so one can argue that the expansion does not significantly affect 
the
network dynamics. The great advantage of Minkowski space is that the
network evolution is very easy to simulate numerically: the code is
generally many times shorter than an equivalent code for a Friedmann
background \cite {AlbTur,FRWCodes}, and makes fewer demands
on both RAM and CPU time. We are therefore
able to get much better statistical significance from the data.

We present results from an extensive programme of numerical simulations 
for the two-time
correlators of $\Theta_{00}$ and $U$, and use a simple model to explain
most of the features we observe. It turns out that, to a good 
approximation,
the network can be thought of as consisting of randomly placed segments 
of string with
random velocities. The segments are of length $\xi$
and number density $\xi^{-3}$, thus reproducing the correct behaviour
for the length density.

In this model it is easy to show that the coherence time-scale in a 
Fourier mode
of wavenumber {\em k} is determined by the time segments take to travel 
a
distance $k^{-1}$. The model in fact predicts that the correlations 
between
the energy density  at times $\eta$ and $\eta'$ decrease as
${\rm exp}( -\sixth k^2 \vsq (\eta-\eta')^2)$, where $\bar v$ is the 
r.m.s string
velocity.
Hence we can talk of a coherence time-scale $\eta_c={\sqrt 3}/{\bar v} 
k$,
which is the time over which the correlation function falls to 
$e^{-\half}$
of its equal time value.
Given that we find $\vsq=0.36$, the model predicts $\eta_c\simeq 3/k$.
Our results actually indicate that at high {\em k}, $\eta_c$ decreases 
faster
than $k^{-1}$, but we have some evidence that this behaviour is a 
lattice artifact.

\section{Flat Space Strings}

We use a development of a code written by one of us some time ago 
\cite{SakVil1,SakVil2}.
Initial string realisations are generated using the 
Vachaspati-Vilenkin algorithm \cite{VV}. This mimics the Kibble 
mechanism for the spontaneous
breaking of a scalar U(1) symmetry with a ``Mexican hat''  potential, 
as the 
Universe cools through a thermal  phase transition.

This initial configuration is set up for evolution on a cubic lattice 
with fundamental
lattice side $\delta$. We are free to alter the initial correlation 
length $\xi_0$
in terms of $\delta$ and we may add structure to the network by placing 
a percentage
$P_c$ of  cusps randomly along the string. Cusps are string links which 
are confined
to one lattice site and which move at the speed of light. Adding cusps 
avoids peculiarities
arising from having long straight segments of string in the initial 
conditions.

In Minkowski space the string equations of motion are
\ben
{\bf X}'' = \ddot {\bf X} 
\label {WaveEq}
\een
where  $\dot { }={\pa / \pa\eta}$ and $\ '={\pa / \pa\sigma}$; $\eta$ 
is time
and $\sigma$ is a space-like parameter along the string. If
we regard Minkowski space as a Friedmann-Roberston-Walker Universe
in the limit that the expansion rate goes to zero, then $\eta$ can be 
identified as FRW conformal time.
${\bf X}={\bf X}(\sigma,\eta)$ is a position three-vector which 
satisfies
\ben
{\bf X}'\cdot\dot {\bf X}=0
\label {GaugeCond1}
\een
\ben
{\bf X}'^2+\dot {\bf X}^2=1   
\label {GaugeCond2}
\een
This is a convenient set of constraints, as (\ref {GaugeCond2}) 
ensures that the energy of a segment of string is proportional
to its length in $\sigma$.

We solve this using the Smith-Vilenkin algorithm \cite {SmithVil:alg} 
which is based on the 
exact finite difference solution to (\ref {WaveEq}),
\ben
{\bf X}(\sigma,\eta+\delta) = {\bf X}(\sigma+\delta,\eta) + {\bf 
X}(\sigma-\delta,\eta) - {\bf X}(\sigma,\eta-\delta)
\label {SmithVil}
\een
If the string points are initially defined on the sites of a cubic 
lattice $(N\delta)^3$,
then (\ref {SmithVil}) ensures that they remain on the lattice at time 
steps of $\delta$.
One can see from (\ref {SmithVil}) that stationary elementary segments 
have length
(and hence energy) of $2\delta$ and point in one of six directions.

Because the string points lie on the sites of the lattice, identifying 
crossing events
is easy. When two strings cross, they intercommute with a probability 
which
is set to 1.
Loops with energy greater than or equal to a threshold value
of $E_c$ are allowed to leave the network,
while reconnections are forbidden for loops with energy equal to $E_c$. 
Forbidding 
reconnections allows
energy to leave the network fairly efficiently; otherwise it
takes much longer for the effect of the initial conditions to wear off.
The natural and usual value for $E_c$ is the minimum segment length,
$2\delta$. In fact, most of the energy
in the string network goes into the smallest possible loops.
In some sense, the cusps model the gravitational radiation of a 
realistic string network:
They travel at the speed of light and do not subsequently interact with 
the network.

Each realisation is evolved on a $(64\delta)^3$ or a $(128\delta)^3$ 
lattice with
approximately $7500$ or $60000$ points describing the string network. 
For calculating
quantities like power spectra and two-time correlation functions we 
typically
average over 50 realisations.

\section{Results}

The energy-momentum tensor for a cosmic string at a point {\bf x} is
\ben
\Theta^{\mu\nu}({\bf x})=\int d\sigma (\dot X^\mu \dot X^\nu - X'^\mu 
X'^\nu)
                 \delta^{(3)}({\bf x}-{\bf X}(\sigma,\eta)) 
\label{emx}
\een
It is simple to calculate this on a cubic lattice for our string 
realisations,
and use a Fast Fourier Transform to get a Fourier mode representation. 
Here we want
to measure the power spectra of $\Theta_{00}(k)$ and $U(k)$. These
are calculated by averaging the amplitudes over all Fourier modes  with 
a wavenumber between
{\em k} and {\em k} + $2\pi/\delta$.

\begin{figure}
\centerline{\epsfig{file=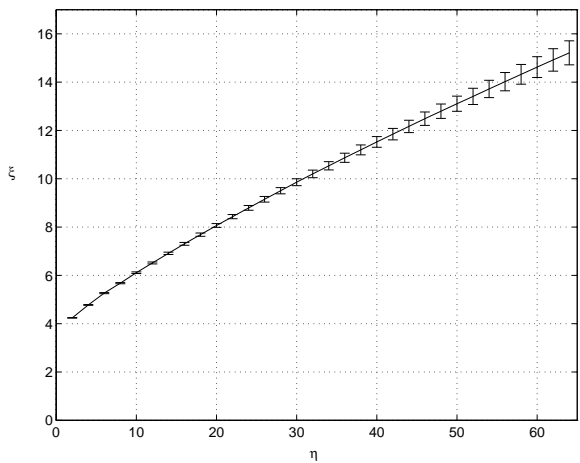,width=3in,angle=0}}
\caption{The energy-density length scale $\xi$ over time $\eta$ for a
$128^3$ lattice, averaged over 50 realisations}
\label{fig:xi}
\end{figure}

Previous work \cite {SakVil1,SakVil2,VHS} has tried to identify 
those features of the network that are scaling, when a relevant length 
scale grows with the 
horizon. For example, we can define the familiar energy density scale 
$\xi$ by $\xi^2=\mu/\rho_{\inf}$ where
$\rho_{\inf}$ is the density of string with energy greater than $\xi$. 
(This
apparently circular definition works because we know the initial step 
length $\xi_0$ and
to calculate $\rho_{\inf}$ we use $\xi$ from the previous time step).
Scaling is reached when $x=\pa \xi/\pa \eta$ is constant.
In the cases considered here $x=0.15$ ($E_c=2\delta$) and $x=0.12$  
($E_c=4\delta$).
The usual definition for scaling, that $x=\xi/\eta$ is
constant, can not be used here as throughout our simulations $\xi_0$ is 
too large to be ignored.
For this reason, we express all scaling functions in terms of $\xi$, 
rather than $\eta$. The
behaviour of $\xi$ is shown in Figure \ref{fig:xi}.

\begin{figure}
\centerline{\epsfig{file=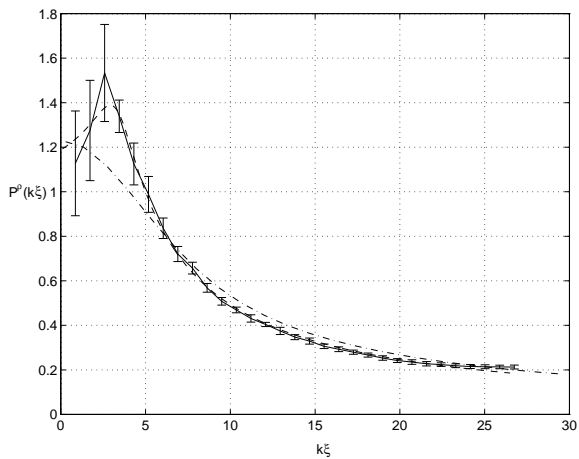,width=3in,angle=0}}
\caption{Scaling function for the energy density as defined by equation 
(\ref{ps00sc1})
The solid line is the measured scaling function with $1-\sigma$ error 
bars from ensemble
averaging. The dashed line is the fitted function equation (\ref{fform}). 
The dash-dot line is the predicted form from our model, the last line
in equation  (\ref{fm3}). }
\label{fig:pseng}
\end{figure}

\begin{figure}
\centerline{\epsfig{file=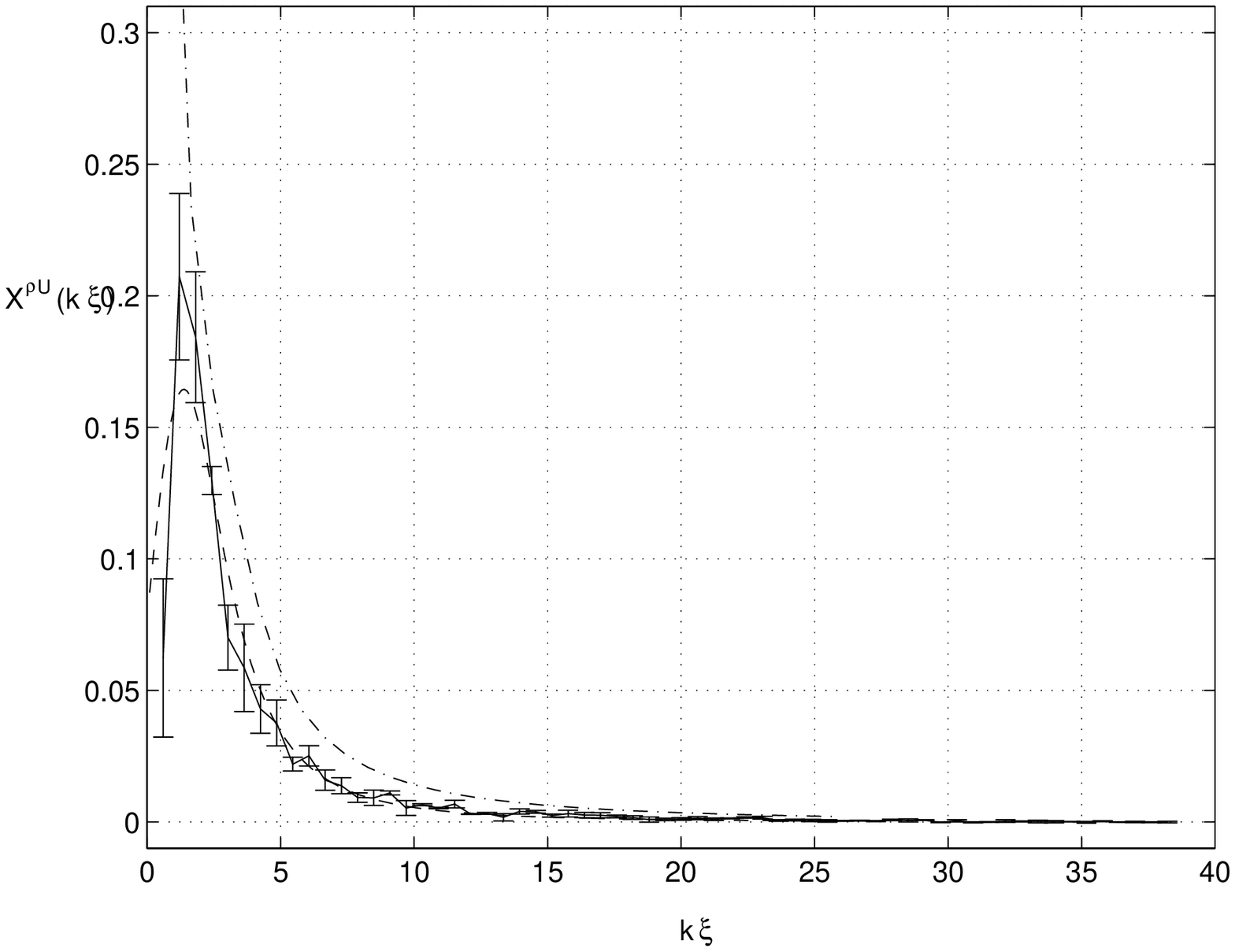,width=3in,angle=0}}
\caption{Scaling fucnction for the equal time energy-momentum cross 
correlator
as defined by equation (\ref {pscrosssc1}). The solid line is the 
measured
scaling function with $1-\sigma$ error bars from ensemble
averaging. The dashed line is the fitted function equation (\ref 
{Xform}).
The dash-dot line is the predicted form from our model,
equation (\ref{Xmodel}).}
\label{fig:pscross}
\end{figure}

\begin{figure}
\centerline{\epsfig{file=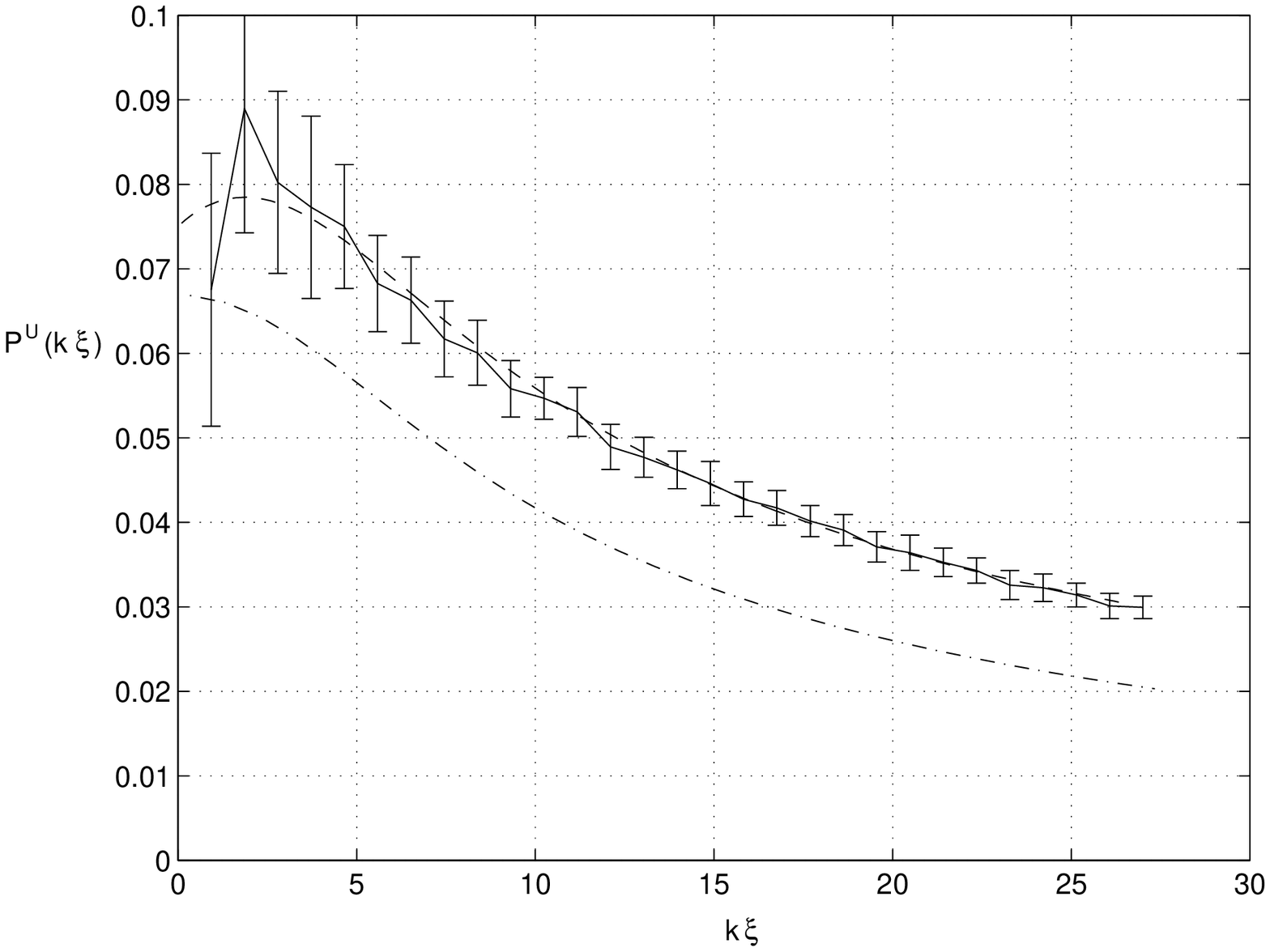,width=3in,angle=0}}
\caption{Scaling fucnction for the momentum density as defined by 
equation (\ref {ps0isc1})
The solid line is the measured scaling function with $1-\sigma$ error 
bars from ensemble
averaging. The dashed line is the fitted function equation (\ref 
{gform}).
The dash-dot line is the predicted form from our model, the last line
in equation  (\ref {momden1}). }
\label{fig:psmom}
\end{figure}

The infinite string energy-momentum power spectra also exhibit scaling 
behaviour, which we 
express in terms of the scaling functions $P^{\rho}$ and $P^{U}$. We 
also consider the equal time
cross correlation function $\pscross$ which is a real function within 
the ensemble errors.
\ben
\psengt = { V P^{\rho}(k\xi) \over \xi }
\label {ps00sc1}
\een
\ben
\pscross = { V X^{\rho U}(k\xi) \over \xi }
\label {pscrosssc1}
\een
\ben
\psmomt = { V P^{U}(k\xi) \over \xi }
\label {ps0isc1}
\een
These scaling forms are fixed by the requirement that the density 
fluctuations
obey the scaling law 
\ben
\int_0^{\xi^{-1}}\,d^3k\, \psengt \propto \xi^{-4}
\label {rms_req}
\een
with a similar law for the other components of $\Theta_{\mu\nu}$.
We display our results for $P^{\rho}$, $X^{\rho U}$  and $P^{U}$ in 
Figures \ref {fig:pseng},
\ref {fig:pscross} and \ref {fig:psmom} along with fits to the 
following forms
\ben
P^{\rho}(z) = { a \over {(1-(bz)+(cz)^n)^{1/n} } }
\label {fform}
\een
\ben
X^{\rho U}(z) = { d \over {(1-(ez)+(fz)^m)^{2/m} } }
\label {Xform}
\een
\ben
P^{U}(z) = { g \over {(1-(hz)+(jz)^p)^{0.66/p} } }
\label {gform}
\een
 which are motivated by a requirement of white noise at large scales 
due to
 uncorrelated initial conditions
 and good single parameter fits to  $P^{\rho} \sim (k\xi)^{-0.98 \pm 
0.06}$,
 $P^{U} \sim (k\xi)^{-0.66 \pm 0.03}$ and  $X^{\rho U} \sim 
(k\xi)^{-2.0 \pm 0.1}$
 at small scales.

 The errors indicate ensemble standard deviation. The other parameters
 are $a=1.18$, $b=0.25$, $c=0.24$, $n=6$, $d=0.23$, $e=0.3$, $f=0.5$,
 $m=3$, $g=0.07$, $h=0.15$, $j=0.18$ and $p=1.59$. 
 We consistently get a peak for $P^{\rho}(z)$ and $P^{U}(z)$ at
 $k\xi \simeq 3$ which corresponds to a physical wavelength of about 
$2\xi$.
 However, it is difficult to be certain about the initial rise in the 
power spectrum
 because the error bars are large. In the notation of Magueijo {\em 
et\, al\,}
 the peak corresponds to an $x_c=k\eta$ of approximately $20$.
  
 As discussed above, the real importance of $\Theta_{00}$ and 
$\Theta_{0i}$ in the context of perturbation
 theory is how they are correlated  over time. We performed $k$-space 
measurements
 of the two-time correlation functions
 \ben
 C^{\rho\rho}(k;\eta,\eta')=V^{-1}\TTeng\\
 \label {TTengdef}
 \een
 \ben
 C^{\rho U}(k;\eta,\eta')=V^{-1}\TTcross
 \label {TTcrossdef}
 \een
\ben
 C^{UU}(k;\eta,\eta')=V^{-1}\TTmom
 \label {TTmomdef}
 \een
 During the scaling era they are well approximated by 
 \ben
 C^{\rho\rho}={1 \over \sqrt {\xi\xi'}} 
\sqrt{P^{\rho}(k\xi)P^{\rho}(k\xi')}
  {\it e}^{-\half \Upsilon^2 k^2 (1-(k\Delta)) (\eta-\eta')^2}
 \label{2t00approx}
 \een
 \ben
 C^{\rho U}={1 \over \sqrt {\xi\xi'}} 
\sqrt{P^{\rho}(k\xi)P^{\rho}(k\xi')}
 {\it e}^{-\half \Upsilon'^2 k^2 (\eta-\eta')^2} ({\alpha \over 
k\sqrt{\xi\xi'}}-\Upsilon'^2 k (\eta-\eta'))
 \label{2tcrossapprox}
 \een
 \ben
 C^{UU}={1 \over \sqrt {\xi\xi'}} \sqrt{P^{U}(k\xi)P^{U}(k\xi')}
 {\it e}^{-\half \Upsilon''^2 k^2 (\eta-\eta')^2} (1-\Upsilon''^2 k^2 
(\eta-\eta')^2)
 \label{2t0iapprox}
 \een
 where we have used the scaling forms of the power spectra and 
cross-correlator
 and $\xi=\xi(\eta)$, $\xi'=\xi(\eta')$. The imaginary component of 
these correlators is consistent
 with zero within the ensemble errors. The forms of these functions are 
motivated by a model
 which we describe in the next section.
 
\begin{figure}
\centerline{\epsfig{file=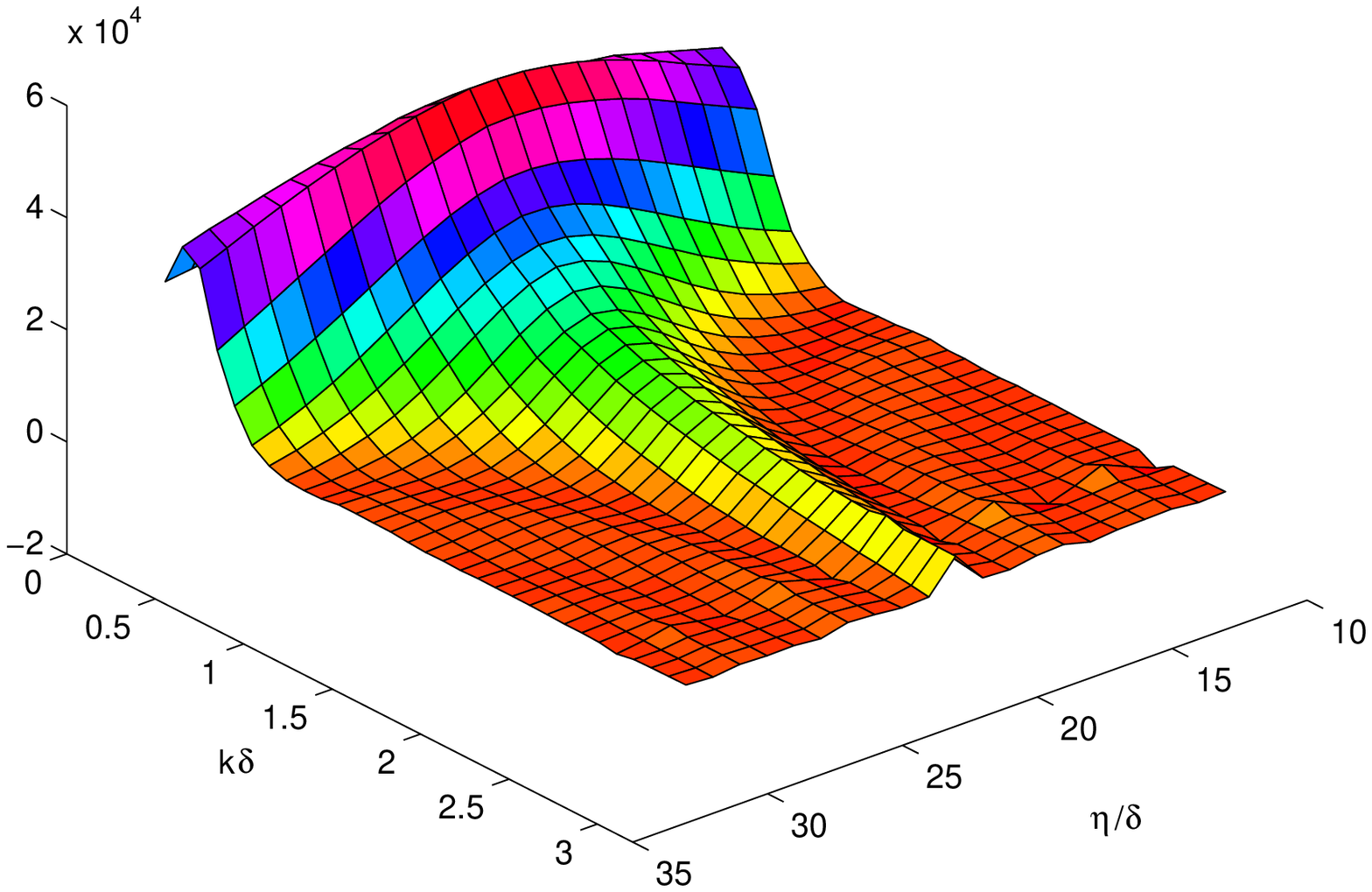,width=3in,angle=0}}
\caption{Two-time correlation function $\TTeng$ for $\eta'=22\delta$}
\label{fig:ttcf00}
\end{figure}

\begin{figure}
\centerline{\epsfig{file=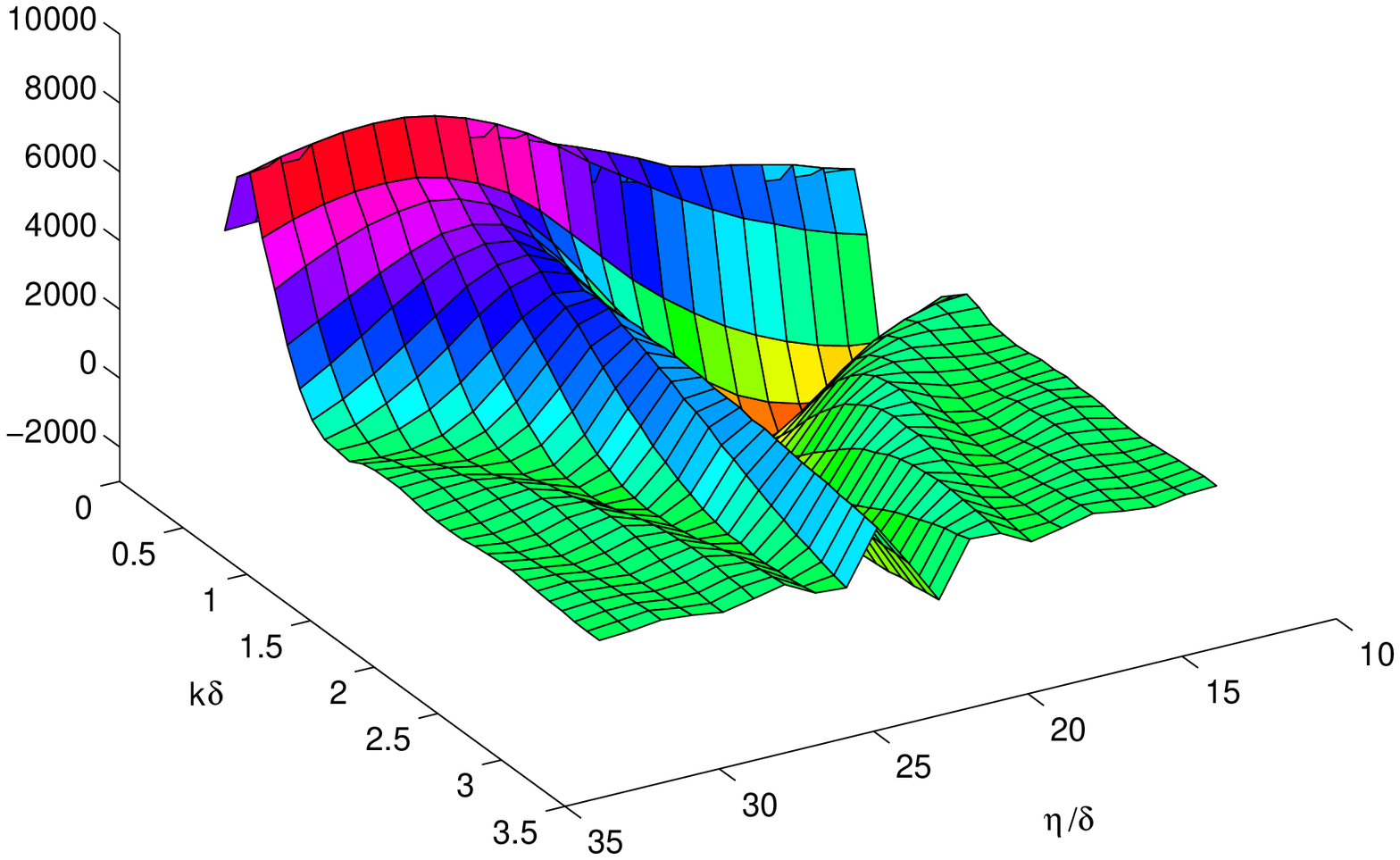,width=3in,angle=0}}
\caption{Two-time correlation function $\TTcross$ for $\eta'=22\delta$}
\label{fig:ttcfcross}
\end{figure}

\begin{figure}
\centerline{\epsfig{file=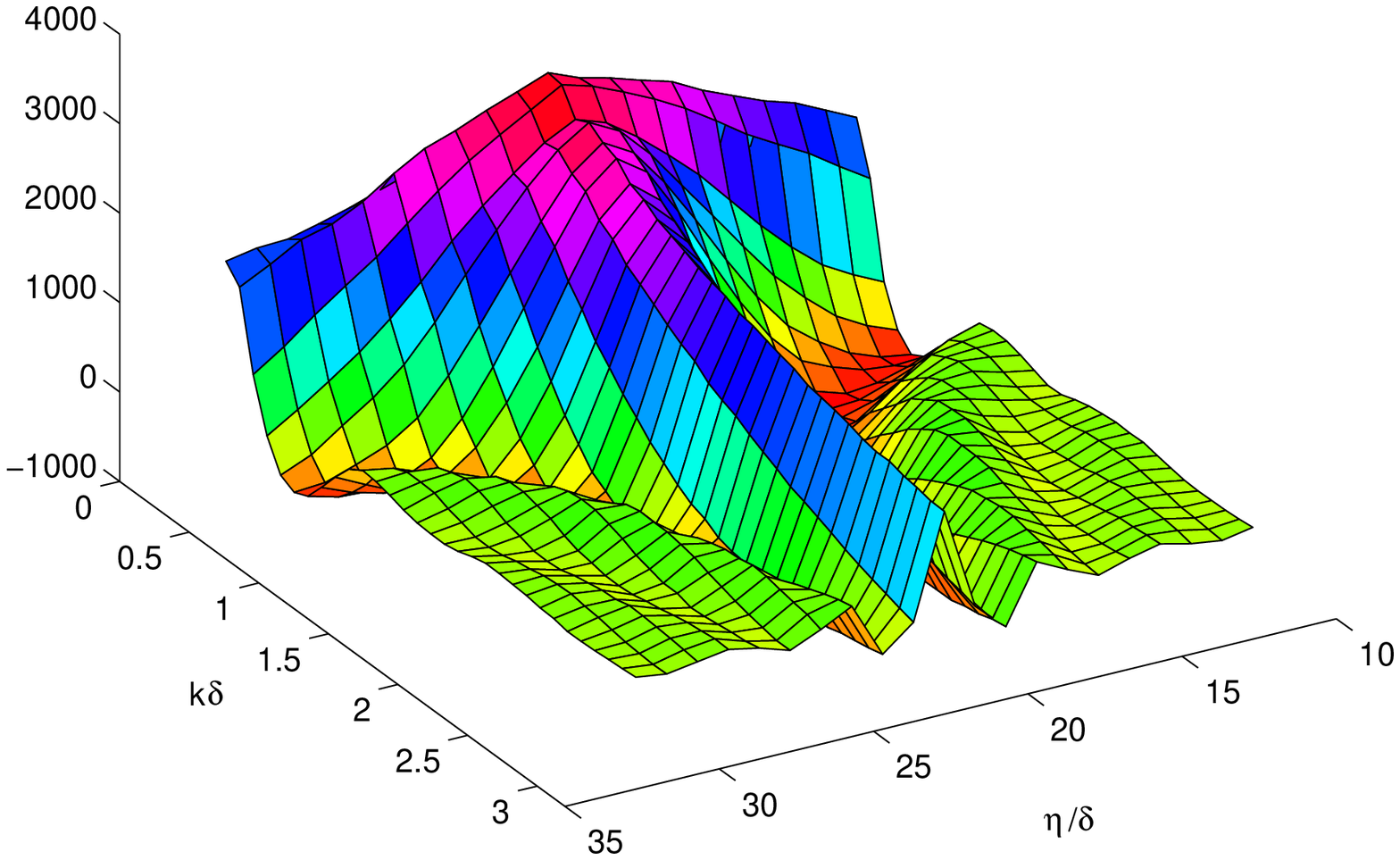,width=3in,angle=0}}
\caption{Two-time correlation function $\TTmom$ for $\eta'=22\delta$}
\label{fig:ttcf0i}
\end{figure}

 The values for $\Upsilon$, $\Upsilon'$ and $\Upsilon''$  are given in 
Table \ref{tab1}.
 $\Delta$ is approximately $3\delta$.  These parameters were determined 
by minimisation of
 the $\chi$-squared for each realisation. The normal distribution of 
parameters
 obtained allows an estimate to be made of each parameter and its 
$1-\sigma$ errors.
 We stress that we consider equations (\ref{2t00approx}), 
(\ref{2tcrossapprox})
 and (\ref{2t0iapprox}) to be an approximation, although a good one for
 $|\eta-\eta'|\,\leq 8 k^{-1}$. For comparison, the energy and momentum
 correlators fall to half their maximum value at $|\eta-\eta'|\,\simeq 
3 k^{-1}$.
 Outside the range given there are two effects
 present in the measured correlators which invalidate
 the model. The first is small $k$-dependent oscillations about zero as
 predicted by Turok \cite{Turok}. The second is sharp peaks in the 
correlators
 at small scales which we take to be an effect of defining
 strings on the lattice. 
\begin{table}[ht]
\begin {center}
\begin {tabular} {|c|c|c|c|c|}
\hline
$E_c$ & $\Upsilon$ & $\alpha$ & $\Upsilon'$& $\Upsilon''$\\
\hline
$2\delta$ & $0.21 \pm 0.05$ & $0.19 \pm 0.05$ & $0.42 \pm 0.05 $ & 
$0.36 \pm 0.07$\\
$4\delta$ & $0.18 \pm  0.05$  & $0.16 \pm 0.05$ & $0.42 \pm 0.05 $ & 
$0.42 \pm 0.06$\\
\hline
\end {tabular}
\caption{Fitted parameters for the models in equations 
(\ref{2t00approx}),
(\ref{2tcrossapprox})  and (\ref{2t0iapprox})}
\label{tab1}
\end{center}
\end{table}
 Figures \ref {fig:ttcf00}, \ref {fig:ttcfcross} and
 \ref {fig:ttcf0i} show the measured functions at a time 
$\eta'=22\delta$ (which
 is within  the scaling era of our simulations). 

 From the gaussian fall-off over time we see that for a given mode with 
wavenumber {\em k}  
 the network energy density decorrelates on a characteristic time scale 
which goes like
 $1/\Upsilon k \sqrt{(1-(k\Delta))}$ and the momentum density varies as
 $1/\Upsilon''k$ for $E_c=2\delta$.
 
\section{Theoretical Model}

\subsection{Power Spectra}

We can use a very simple model to account for the forms given in (\ref 
{fform})
and (\ref{gform}), if not the precise details. From (\ref {emx}) we get
\ben
\Theta^{\mu\nu}({\bf k})=\int d\sigma\,(\dot X^\mu \dot X^\nu - X'^\mu 
X'^\nu)
                     e^{i {\bf k}\cdot{\bf X}(\sigma,\eta)}
\label {emk} 
\een
At any given time the power spectrum is 
\ben
\psengt=\la {\int d\sigma\, d\sigma'\, {\it e}^ {i {\bf k}\cdot({\bf 
X}(\sigma,\eta)-{\bf X}(\sigma',\eta))} \ra}
\label {fm1}
\een
We then make some assumptions about the statistics of the network. 
Firstly, that for
each lag $\sigma_-=\sigma-\sigma'$, the quantities 
$X_i(\sigma)-X_i(\sigma')$,\,$\dot X_i(\sigma)$ and $X'_i(\sigma)$
are gaussian random variables with zero mean. Secondly, that the 
distribution of strings
is isotropic. Then
\begin{eqnarray}
\pseng & = & \int d\sigma\, d\sigma'\, {\it e}^ {-\sixth k^2 \la{({\bf 
X}(\sigma)-
{\bf X}(\sigma'))^2\ra}}\nonumber\\[-3mm]\\
& = & \half\int d\sigma_+\, d\sigma_-\, {\it e}^ {-\sixth k^2 
\Gamma(\sigma_-)}\nonumber
\label{fm2}
\end{eqnarray}
where we have introduced the function $\Gamma(\sigma-\sigma')=\la{({\bf 
X}(\sigma)-{\bf X}(\sigma'))^2\ra}$
and changed variables to $\sigma_-=\sigma-\sigma'$ and 
$\sigma_+=\sigma+\sigma'$.
We will also use $\tsq=\tansq$. The third assumption is that the 
network is described by a collection
of randomly placed string segments of length $\xi/{\bar t}$ and 
total energy ${V  \xi^{-2}}$, where $V$ is the volume of the simulation 
box.
Then $\Gamma=\tsq \xi^2 ({ \sigma_-/\xi } )^2$ and $L=\half\int 
d\sigma_+  = {V  \xi^{-2}}$.
If we define $z=\sigma_-\bar t /\xi$ then
\begin{eqnarray}
\psengt & =& L \int d\sigma_-\, {\it e}^ {-\sixth k^2 \tsq \xi^2 
(\sigma_-/\xi)^2}\nonumber\\
& = & {L\xi \over {\bar t} } \int_{-\half}^{\half} dz\, {\it e}^ 
{-\sixth (k\xi)^2z^2 }\label {fm3}\\
& = & { V \over \xi {\bar t} }{2 \sqrt{6} \over k\xi } {\rm{erf}} 
({k\xi \over 2 \sqrt{6}})\nonumber
\end{eqnarray}
Thus the scaling form of (\ref {ps00sc1}) emerges quite naturally.
If we compare (\ref {fm3}) with (\ref {ps00sc1}), we can write the 
predicted form of the
scaling function $P^{\rho}(k\xi)$. The large and short wavelength 
limits are
\ben
P^{\rho}(k\xi) = \left\{ \begin{array}{ll}
              (k\xi)^0 & {\rm if} \: k\xi << 2\sqrt{6} \\
              (k\xi)^{-1} & {\rm if} \: k\xi >> 2\sqrt{6} \\
            \end{array}
           \right.
\label{gb}
\een
This function is plotted along with the measured and fitted model in 
Figure \ref {fig:pseng}.
It should be noted that the normalisation is given by the model along 
with the
measurement for $\bar t$, and not introduced
by hand. This model does better at predicting the measured
power spectrum than a random walk model, in which the power
on large scales goes like $k^{-2}$. Therefore, although each individual 
string is a random walk,
they must be highly (anti) correlated on large scales.

Similarly, the power spectrum for $U$ becomes
\begin{eqnarray}
\psmom & = & \la {\int d\sigma\, d\sigma'\, \hat k_i\hat k_j \dot 
X^i(\sigma) \dot X^j(\sigma')
{\it e}^ {i {\bf k}\cdot({\bf X}(\sigma,\eta)-{\bf X}(\sigma',\eta))} 
\ra}\nonumber\\
 & = &  L \int d\sigma_-\, (\tthird {\cal V}(\sigma_-) - \tninth k^2 
\Pi^2(\sigma_-))
{\it e}^ {-\sixth k^2 \Gamma(\sigma_-)}\label{momden1}\\
 & = & {2 V \over 3 \xi {\bar t} } \int_{0}^{\half} dz\,
 (\tilde {\cal V}(z/\bar t) - \tthird (k\xi)^2 \tilde\Pi^2(z/\bar t)
{\it e}^ {-\sixth (k\xi)^2 z^2 }\nonumber
\end{eqnarray}
where ${\cal V}$ and $\Pi$ are the correlation functions
\ben
{\cal V}(\sigma)=\la{ {\bf\dot X}(\sigma)\cdot{\bf\dot X}(0)\ra}
\label{correlV}
\een
\ben
\Pi(\sigma)=\la{ {\bf\dot X}(\sigma)\cdot({\bf X}(\sigma)-{\bf X}(0) 
)\ra}
\label{correlPI}
\een
and $\tilde {\cal V}(\sigma_-/\xi)$ and  $\tilde\Pi(\sigma_-/\xi)$ are 
their scaling forms.
We have measured these correlation functions elsewhere \cite{VHS},
and we can use them to calculate the scaling function $P^{U}$. The
result is plotted in Figure \ref {fig:psmom}. Although for very large 
$k\xi$
this model predicts a $(k\xi)^{-1}$ behaviour, within the $k$-range of 
our simulations
the model agrees well with our measurements giving a fit to 
$(k\xi)^{-0.66}$, although
the normalisation is not so impressive as for the energy-density.
\vskip 0.1in
\subsection{Two-time correlation functions}

\begin{figure}
\noindent
\centering
\centerline{\epsfig{file=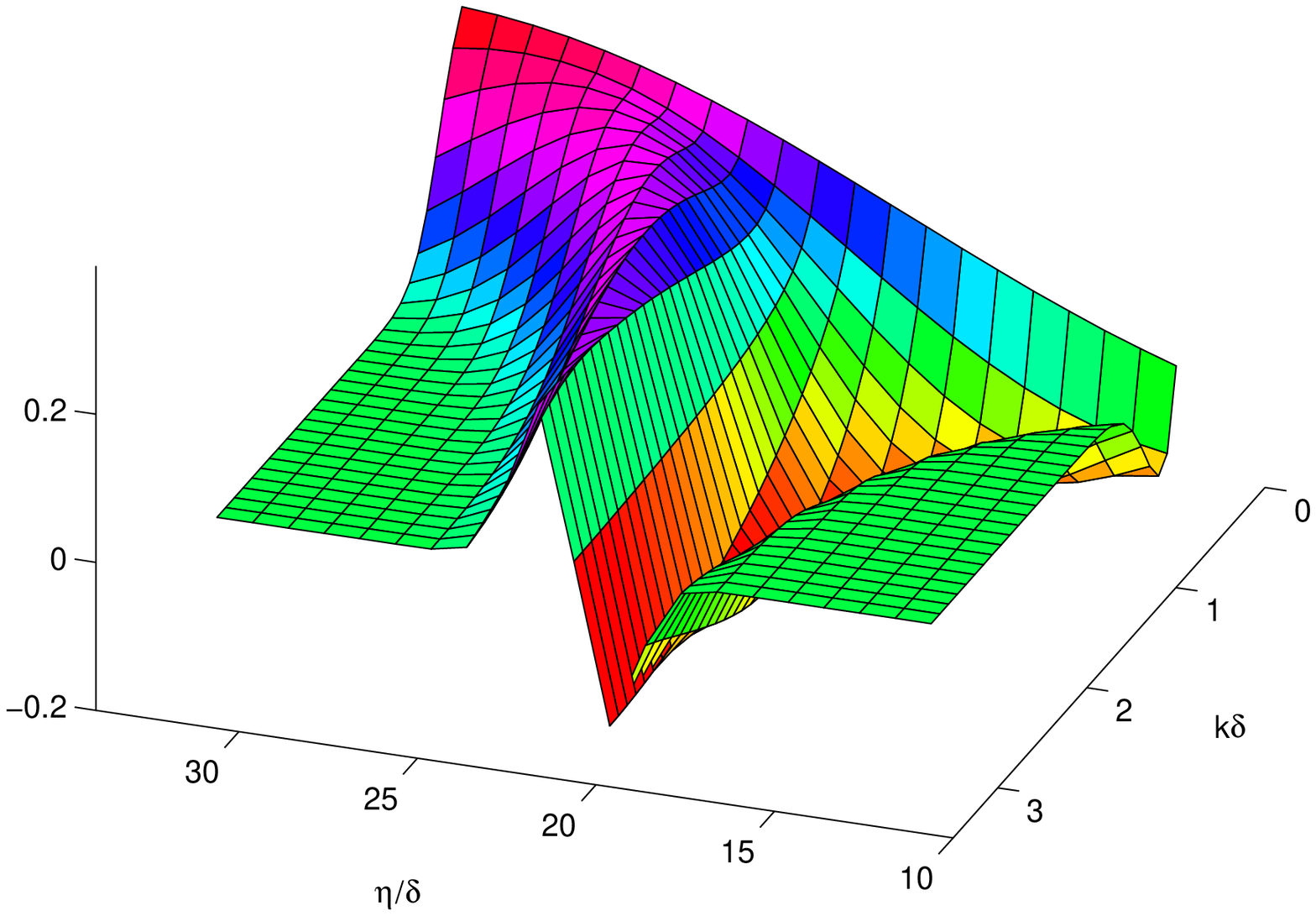,width=3in,angle=0}}
\caption{Model of time dependence of $\TTcross$}
\label{fig:ttcf_model_timecross}
\end{figure}

\begin{figure}
\noindent
\centering
\centerline{\epsfig{file=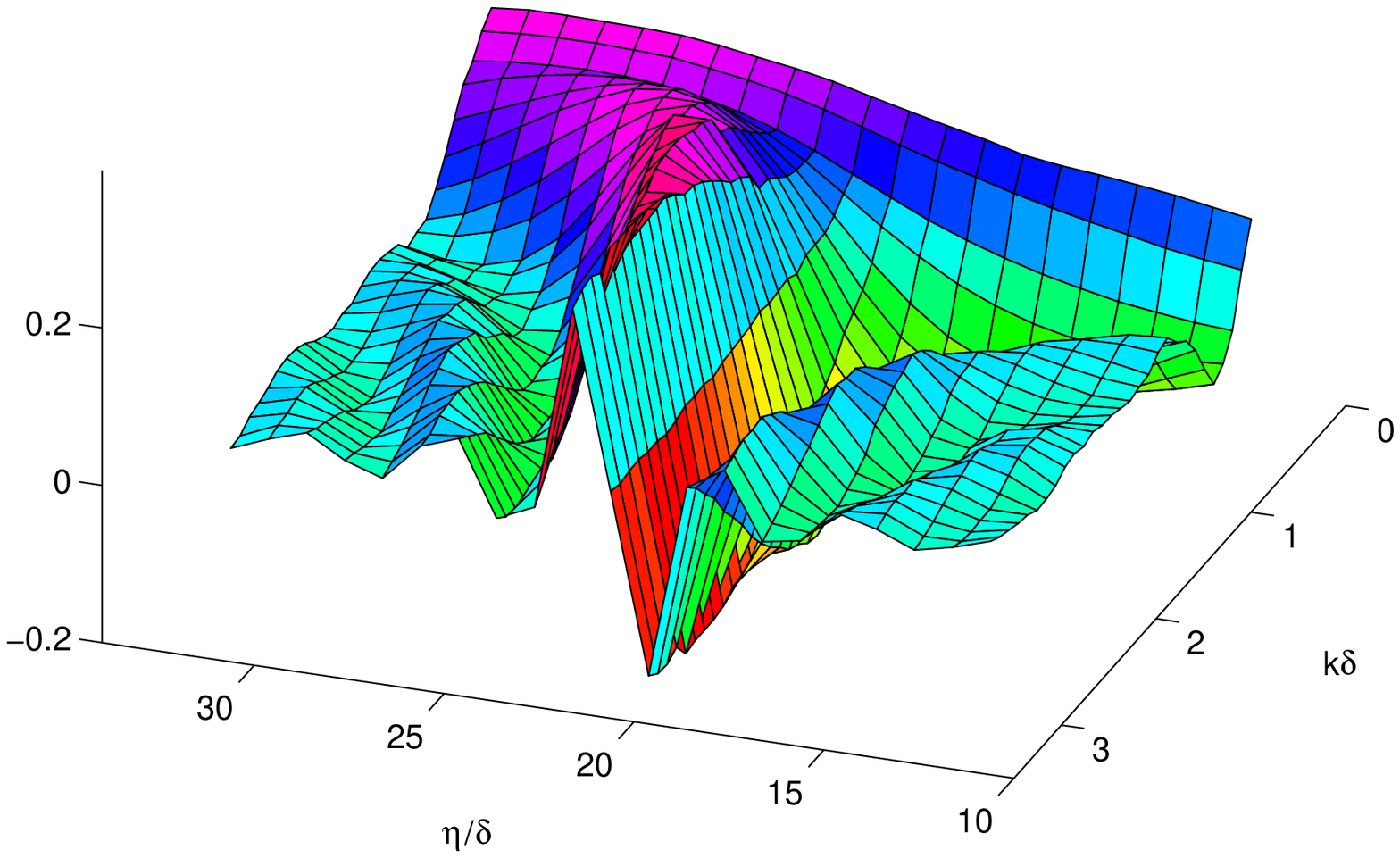,width=3in,angle=0}}
\caption{Measured time dependence of $\TTcross$}
\label{fig:ttcf_timecross}
\end{figure}

In order to calculate the two-time correlation functions in this 
framework we need
the two-time correlation functions $\Gamma(\sigma,\sigma',\eta,\eta')$,
${\cal V}(\sigma,\sigma',\eta,\eta')$ and 
$\Pi(\sigma,\sigma',\eta,\eta')$. However, we have not yet
measured these quantities and instead resort to crude modelling. If we 
consider small
scales  we may assume that each segment is moving at a velocity 
$\vsq=\velsq$ in a
random direction orthogonal to the orientation of the segment so that 
\ben
\Gamma(\sigma_-,\eta,\eta')=\tsq\sigma_-^2 + \vsq (\eta-\eta')^2
\label {gam1}
\een
Using equations (\ref {fm2}) and (\ref {gam1}) we get 
\ben
\TTeng=\psengt {\it e}^{-\sixth \vsq k^2(\eta-\eta')^2}
\label {TTmod2}
\een
However this model does not reflect the fact that between $\eta$ and 
$\eta'$ some energy
is lost and the integrand is no longer independent of $d\sigma_+$. We 
know
that $C^{\rho\rho}(\eta,\eta')=C^{\rho\rho}(\eta',\eta)$ and the most 
natural way of respecting this condition is
to replace the power spectrum at $\eta$ with the square root of the 
product of the 
power spectra at the two times. Hence the final result to be compared 
with equation
(\ref {2t00approx}) is
\ben
\TTeng={1 \over \sqrt {\xi\xi'}} 
\sqrt{P^{\rho}(k\xi)P^{\rho}(k\xi')}\,{\it e}^{-\sixth \vsq 
k^2(\eta-\eta')^2}
\label {TTmod3}
\een
To model the other two-time correlators $\TTmom$ and $\TTcross$ we must 
be
careful about the conservation of energy-momentum as loops are created 
and energy
is lost from the long string network. If we assume that the loop 
production occurs
evenly along the string and that we are in a scaling regime we can 
model in $k$-space  
the rate of energy going into loops as
\ben
\Lambda(k,\eta)={\lambda(k\xi) \over \xi}\Theta_{00}(k,\eta)
\label {Lambda}
\een
The energy-momentum conservation equation becomes
\ben
U(k,\eta)={1 \over k}{\pa \over \pa \eta}\Theta_{00}+{\lambda \over k 
\xi}\Theta_{00}(k,\eta)
\label {em1}
\een
and hence,
\begin {eqnarray} 
\TTcross & = & \int d\sigma\, d\sigma'\ ({1 \over k}{\pa \over \pa 
\eta}
          +{\lambda \over  k \xi}){\it e}^{-\sixth k^2 (\vsq 
(\eta-\eta')^2 + \tsq\sigma_-^2)}\nonumber\\
& = & \int d\sigma\, d\sigma'\ ({\lambda \over  k \xi} -\tthird k 
\vsq\eta_-))
          {\it e}^{-\sixth k^2 (\vsq (\eta-\eta')^2 + 
\tsq\sigma_-^2)}\label{TTcrossmod1}\\
& = & \psengt ({\lambda \over k\xi}-\tthird k \vsq\eta_-)
{\it e}^{-\sixth \vsq k^2(\eta-\eta')^2}\nonumber
\end {eqnarray} 
To ensure that we do not pick out a time $\eta$ we express $C^{\rho 
U}(\eta,\eta')$
\ben
\TTcross=\TTeng\,({\lambda \over k\sqrt{(\xi\xi')}}-\third k 
\vsq\eta_-)
\label{TTcrossmod3}
\een
The time dependence of this function is  plotted in Figure
\ref {fig:ttcf_model_timecross}, and should be compared with
Figure \ref {fig:ttcf_timecross}.

These plots assume a 
value of $\lambda$ measured from the simulations.
We find it to be roughly constant at large 
$k\xi$ at $\lambda\simeq 0.32\,\, (\pm 0.03)$. Including loop 
production
shifts where the cross-correlator goes to zero from $\eta=\eta'$ to  
$\lambda-\third k^2 \vsq\eta_-\sqrt{(\xi\xi')})=0$.
The effect of energy loss through
loop production on the momentum of the long string may also account for
the form of the equal time cross-correlator.
If we take $\eta=\eta'$ in equation (\ref {TTcrossmod3}), which in 
terms
of this model gives the correlation between the long string
energy and the reaction momentum from loop production, we get
\ben
X^{\rho U}(k\xi)={\lambda \over k\xi}P^{\rho}(k\xi)
\label{Xmodel}
\een
This neatly explains the $(k\xi)^{-2}$ dependence in equation 
(\ref{Xform}),
although the measured normalisation is only $60\%$ of that given by the
independently measured $\lambda$ and equation (\ref{Xmodel}). One 
reason for this
discrepancy may be that our model assumes loop production occurs evenly 
along the string
at a constant rate,
whereas we observe loop production in more discrete bursts.
The prediction from equation (\ref{Xmodel}) is plotted in Figure
\ref {fig:pscross}, along with the measured scaling function.

\begin{figure}
\centerline{\epsfig{file=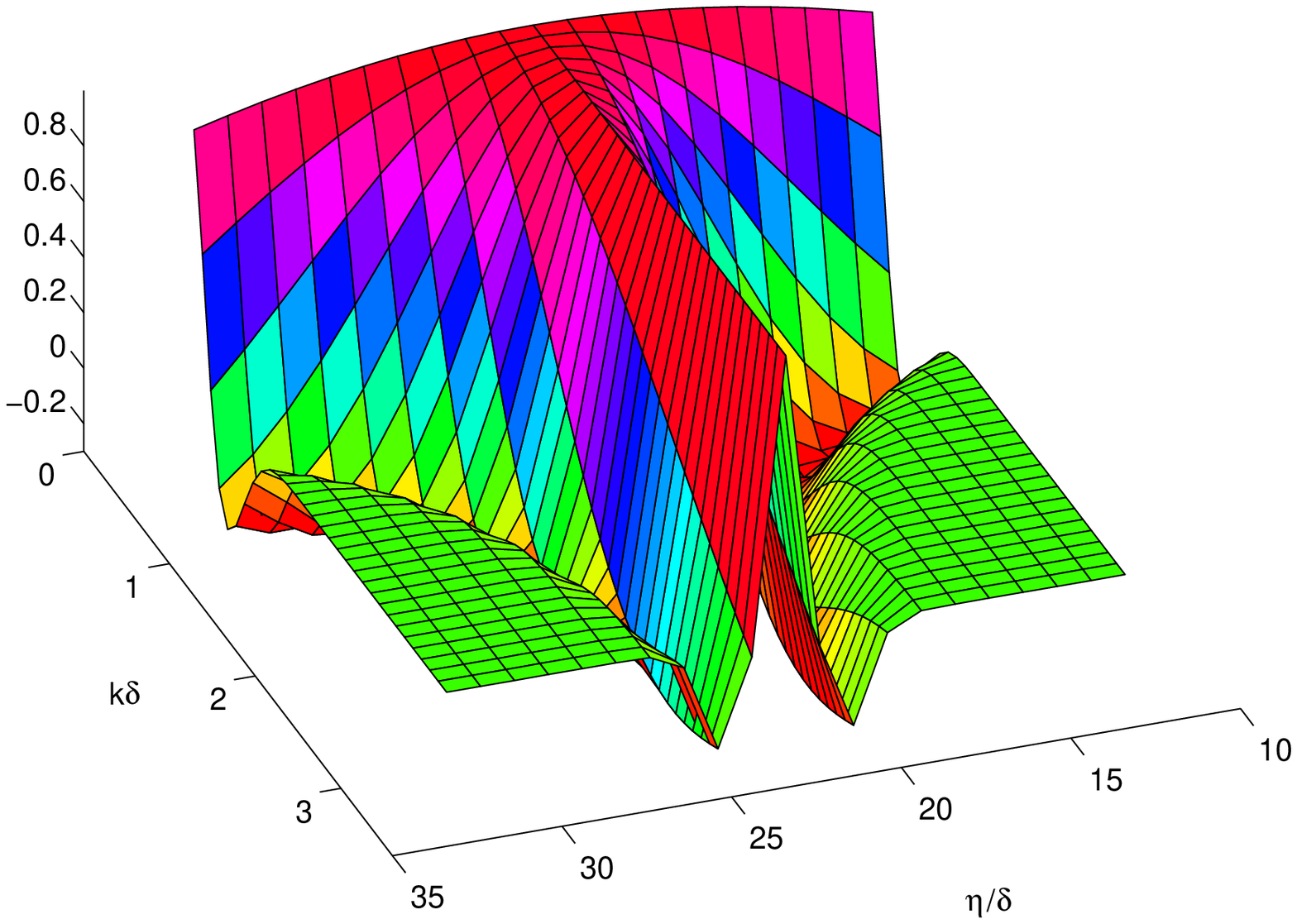,width=3in,angle=0}}
\caption{Model of time dependence of $\TTmom$ }
\label{fig:ttcf_model_time0i}
\end{figure}

\begin{figure}
\centerline{\epsfig{file=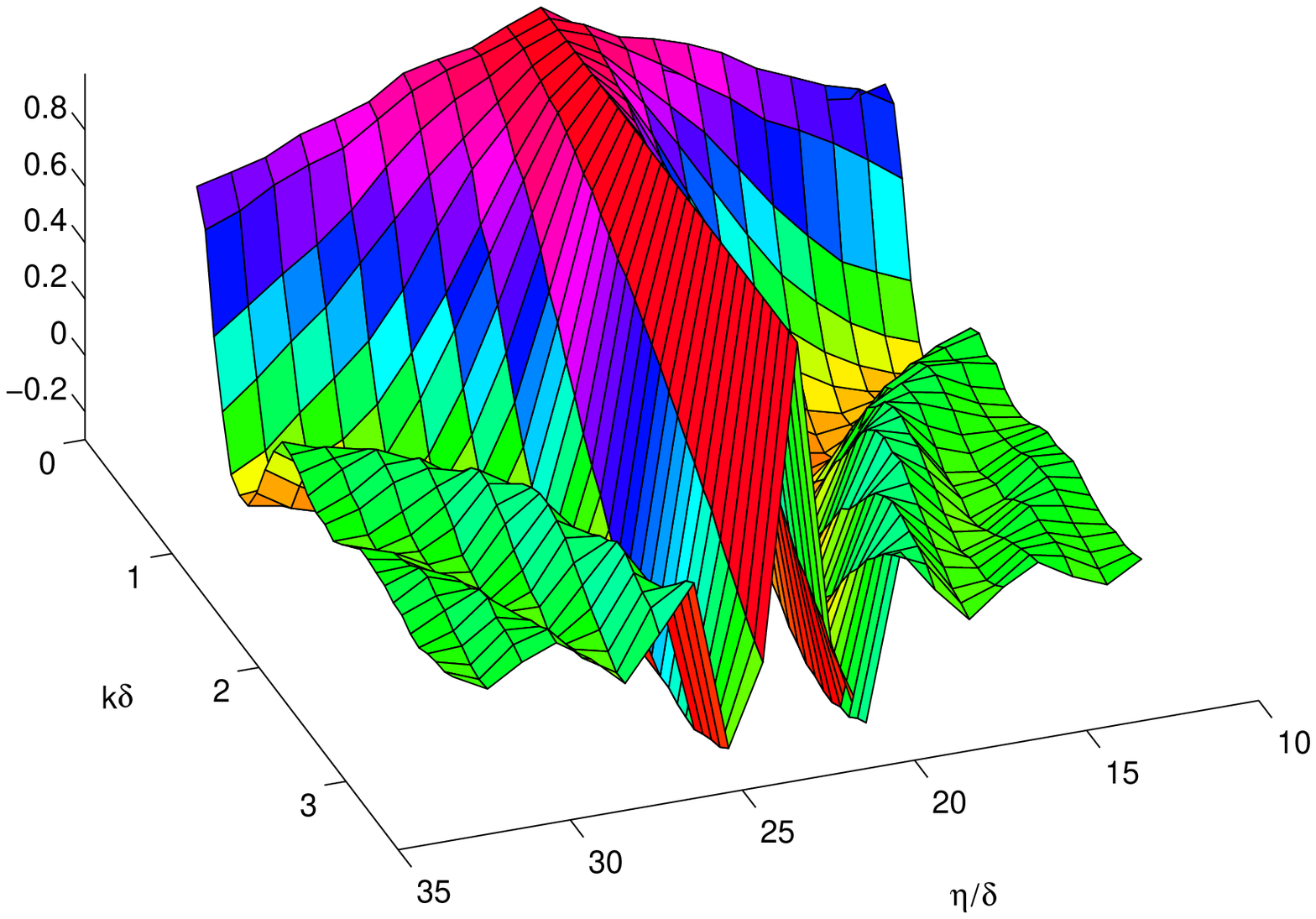,width=3in,angle=0}}
\caption{Measured time dependence of $\TTmom$}
\label{fig:ttcf_time0i}
\end{figure}

We find that the effect of the loop production term is insignificant
for $\TTmom$ and we will drop it in the following expression:
\begin {eqnarray} 
\TTmom & = & {1 \over k^2} \int d\sigma\, d\sigma'\ {\pa \over \pa 
\eta}{\pa \over \pa \eta'} 
          {\it e}^{-\sixth k^2 (\vsq (\eta-\eta')^2 + 
\tsq\sigma_-^2)}\nonumber\\[-3mm]\\
& = & \psengt {\vsq \over 3} (1-{\vsq \over 3} k^2 (\eta-\eta')^2)
    {\it e}^{-\sixth\vsq k^2 (\eta-\eta')^2} \nonumber
\label{TTmommod}
\end {eqnarray} 
Again to ensure $C^{UU}(\eta,\eta')=C^{UU}(\eta',\eta)$, we use the 
product of the square root of the power
spectra, so that 
\ben 
\TTmom = {1 \over \sqrt {\xi\xi'}} 
\sqrt{P^{\rho}(k\xi)P^{\rho}(k\xi')}\,
 {\vsq \over 3} (1-{\vsq \over 3} k^2 (\eta-\eta')^2) {\it 
e}^{-\sixth\vsq k^2 (\eta-\eta')^2} 
\label{TTmommod2}
\een
Equation (\ref{TTmommod2}) incorrectly predicts $U$ to have the same
$k\xi$ dependence in its power spectrum as the energy, with a relative
normalisation given by $\vsq/3$.
We expect a proper treatment in terms
of the two-time on-string correlators for $\Pi$, ${\cal V}$ and 
$\Gamma$
to remedy this problem. The time dependence of this function is plotted 
in 
Figure \ref {fig:ttcf_model_time0i}. As one can verify by comparison 
with 
Figure \ref {fig:ttcf_time0i}, the simplest of assumptions has provided
a reasonable approximation to the measured time dependence.

In this section we have motivated the forms in equations 
(\ref{2t00approx}),
(\ref{2tcrossapprox}) and (\ref{2t0iapprox}). From the model, and using
the measurements $\tsq=0.637 \pm 0.010$ and $\vsq=0.363 \pm 0.010$,
we can make predictions for the coherence parameters $\Upsilon$, 
$\Upsilon'$
and $\Upsilon''$, and the relative normalisation $R=\TTeng / \TTmom$.
The measured values are for $E_c=2\delta$.
\begin{table}[ht]
\begin {center}
\begin {tabular} {|c|c|c|}
\hline
quantity & model  & measured\\
\hline
$\Upsilon$ & $0.35 \pm 0.02$ &  $0.21 \pm 0.05$\\
$\Upsilon'$ & $0.35 \pm 0.02$ &  $0.42 \pm 0.05$\\
$\Upsilon''$ & $0.35 \pm 0.02$ &  $0.36 \pm 0.07$\\
$R$ & $0.35 \pm 0.02$ & $0.38 \pm 0.1$\\
\hline
\end {tabular}
\caption{Parameters for the forms given in equations 
(\ref{2t00approx}), (\ref{2tcrossapprox})
and (\ref{2t0iapprox}) as predicted by our model and measured from the 
simulations. Note
a complication in comparing the values of $\Upsilon$, due to a lattice 
effect in
equation (\ref{2t00approx}) as discussed in the text.}
\end {center}
\end{table}
The comparison for $\Upsilon$ is complicated by an
apparent lattice effect accounted for in equation (\ref{2t00approx}) by
the length scale $\Delta$.
We assume that as $\Delta$ goes to zero and the form in
equation (\ref{2t00approx}) approaches that in equation (\ref{TTmod3}),
allowing a comparison to be made.
Not surprisingly, the model is not wholly satisfactory. We know that
the momentum density does not have the same $(k\xi)$ dependence as the 
energy density,
which is predicted by (\ref {TTmommod2}). Also from (\ref {2t00approx}) 
we see that the strings
decorrelate on a time scale  $\sim 1/k\sqrt{(1-(k\Delta))}$ and not 
$k^{-1}$.
One might think that both features are understandable as it is well 
known that strings move faster 
on smaller scales. This gives faster decorrelation and greater power
in the momentum density on small scales than the simple model would 
suggest. 
Having said this, we note that the scale $\Delta$ introduced in 
equation
(\ref {2t00approx}) is close to the lattice scale and is fairly
constant throughout the simulation. Consequently,
the departure from a $k^{-1}$ coherence time may be due to the fact 
that the strings
are defined at all times on the lattice. It is puzzling that this
effect does not show up in the other correlators. Furthermore, the 
model does not account for the 
oscillations about zero in the two-time correlators. The oscillations 
are
present in all three two-time correlators, but are most obvious
in Figures \ref {fig:ttcf_timecross} and \ref {fig:ttcf_time0i}. As 
Turok has recently pointed out \cite {Turok}, such oscillations should
appear in the Fourier transforms of two-time spatial correlators 
because of a 
causality constraint which sets the correlator to zero for causally
disconnected points.

\section{Conclusions}

We have demonstrated the scaling properties of the power
spectra and cross correlator of two important energy-momentum 
quantities, the 
density $\Theta_{00}$ and the velocity $U$, and
given the large {\em k} behaviour.  We have also studied the two-time
correlators and measured the time coherence in the network.

We find that the energy and the momentum power spectra are peaked at 
around
$k\xi\simeq 3$, where $\xi\simeq 0.15\eta$, thereafter decaying as
$(k\xi)^{-1}$ and $(k\xi)^{-0.66}$ respectively over the range
of our simulations. The cross-correlator decays as $(k\xi)^{-2}$ and is 
peaked at
$k\xi\simeq 2$.

For a mode of spatial frequency $k$, the two-time correlation functions 
of $\Theta_{00}$
and $U$ display correlations over time scales of $\eta_c\simeq 
4.7k^{-1}$
and $\eta_c\simeq 2.4k^{-1}$ respectively. The time scale $\eta_c^2$ is 
the variance
in the Gaussian fall-off as a function of the time difference (see 
equations
(\ref{2t00approx}) and (\ref{2t0iapprox})).
We have presented simple models to explain the qualitative results,
predicting the power spectra and cross correlator including 
normalisations,
to a reasonable accuracy,
and accounted for the features of the two-time correlation functions.

The model describes the string network as a set of randomly placed
segments of length $\xi/\bar t$, where $\bar t=(1-\vsq)^{\half}$,
with random velocities. We assume that relevant
quantities such as velocities and extensions between points with a 
given separation
in $\sigma$ are gaussian random variables. We can then reduce ensemble 
averaging to the study of
two-point correlations along the string, which we must model or 
measure.

From the model we can show that the characteristic coherence time
scale for a mode of spatial frequency $k$ is $\eta_c\simeq 3/k$.
Our simulations give a similar result, although at very small scales
$\eta_c$ decrease faster than $k^{-1}$. This we believe to be a
lattice effect.

There are potential pitfalls in taking the Minkowski space string 
network
as a source for the fluid perturbation variables in a Friedmann model. 
For example,
the energy conservation equation is modified to
\ben
\dot\Theta_{00} + {\dot a \over a}(\Theta_{00}+\Theta)-kU = -\Lambda
\label {emconFRW}
\een
where $a(\eta)$ is the scale factor and $\Theta=\Theta_{ii}$ the trace
of the spatial components of $\Theta_{\mu\nu}$. Since $\Theta$ is
unconstrained by any conservation equation, its fluctuations could 
drive a
fluctuating component in $\Theta_{00}$ whose time scale would go like
${a /\dot a}$ \cite {Stebbins}. However, we find that
${\la |\Theta|^2\ra}<<{\la |\Theta_{00}|^2\ra}$ and believe
the introduction of the $\dot a / a$ term to be a small effect.

If we take the simulations at face value, the implications for the
appearance of the Doppler peaks are not entirely clear cut. The
coherence time is smaller than, but of the same order of magnitude as,
the period of acoustic oscillations in the photon baryon fluid at
decoupling, which is roughly $11/k$ \cite{Peeb}. This is in turn 
smaller 
than the time at which the power in the energy and velocity sources 
peak,
approximately $20/k$. The computations of Magueijo {\em et\, al\,}
\cite {MACF2} of the Microwave Background angular power spectrum for
various source models suggest small or absent secondary peaks. However,
our string correlation functions can be
used as realistic sources to settle the issue.

\section*{Acknowledgements}

We wish to thank Andy Albrecht, Nuno Antunes, Pedro Ferreira,  Paul 
Saffin
and Albert Stebbins for useful discussions.

GRV and MBH are supported by PPARC, by studentship number 94313367, 
Advanced
Fellowship number B/93/AF/1642 and grant number GR/K55967. MS is 
supported by the Tomalla Foundation. Partial support is
also obtained from the European Commission under the Human Capital and 
Mobility programme,
contract no. CHRX-CT94-0423.

\end{document}